\documentclass[preprint]{elsarticle}
\pdfoutput=1 
\usepackage[T1]{fontenc} 
\usepackage{amssymb,amsmath,amsfonts}
\usepackage{hyperref}  
\usepackage{lineno}
\usepackage{float}
\usepackage{graphicx}
\usepackage{xcolor}
\usepackage{tensor}
\usepackage[utf8]{inputenc}
\usepackage{caption,subfig}
\usepackage{slashed}
\usepackage{feynmf}	
\usepackage{longtable}
\usepackage{tabu}
\usepackage{tikz}
\usetikzlibrary{trees}
\usetikzlibrary{decorations.pathmorphing}
\usetikzlibrary{decorations.pathreplacing}
\usetikzlibrary{decorations.markings}
\usetikzlibrary{arrows.meta}

\begin{document}
\title{Combinatorial aspects in the one-loop renormalization of higher derivative theories}
\author{Christian F. Steinwachs}
\ead{christian.steinwachs@physik.uni-freiburg.de}
\address{Physikalisches Institut, Albert-Ludwigs-Universit\"at Freiburg, \\Hermann-Herder-Str. 3, 79104 Freiburg, Germany}
\date{\today}
\begin{abstract}
An efficient way to calculate one-loop counterterms within the Feynman diagrammatic approach and dimensional regularization is to expand the propagators in the integrands of the  Feynman integrals around vanishing external momentum. In this way, a generic one-loop diagram is reduced to a sum of vacuum diagrams. The logarithmically divergent part can be extracted by power counting arguments.
In case of higher derivative theories, the standard implementation of this procedure on a computer algebra system can become quickly inefficient due to a high proliferation of terms coming from the intermediate replacement of high-rank tensor-integrals with symmetrized product of metric tensors. In this note we present a simple combinatorial solution to this problem which makes the implementation much more efficient. This method is especially relevant in the renormalization of higher derivative theories, but might as well be integrated as a standard routine in existing computer algebra programs designed to automatize Feynman diagrammatic calculations.
\end{abstract}
\maketitle

\section{Introduction}
The calculation of one-loop ultraviolet (UV) divergences for a generic relativistic local quantum field theory might be considered as solved problem -- at least at the formal level. This encompasses non-renormalizable and effective theories.
However, explicit calculations can become cumbersome, especially in curved spacetime or in higher derivative theories. 
For the calculation of UV divergences in curved spacetime, the combination of the background field method with heat kernel techniques provides a manifest covariant and efficient tool \cite{DeWitt1965, DeWitt1967a, DeWitt1967b, Abbott1982, Barvinsky1985}.
In particular, the generalized Schwinger-DeWitt algorithm reduces the closed-form calculation of one-loop divergences in field theories with higher derivatives and non-minimal fluctuation operators to the evaluation of products of nested commutators and a few `universal functional traces' \cite{Barvinsky1985}.
There are, however, important cases in which these algorithms are either not directly applicable, such as for fluctuation operators with a degenerate principle part \cite{Ruf2018,Ruf2018b,Ruf2018c}, or become practically inefficient, such as in higher derivative theories like e.g. in Galileon models relevant in cosmology  \cite{Nicolis2009,Deffayet2009,Chow2009,Heisenberg2018a}.\footnote{For a given theory defined by a local action functional, the fluctuation operator is defined as the differential operator resulting from the Hessian of this action.}  In both cases, new techniques are required. 
In flat spacetime, Feynman diagrammatic algorithmic one-loop calculations have been essentially developed in \cite{Hooft1979,Passarino1979}. For an overview, which also includes modern approaches to loop calculations, see e.g. the reviews \cite{Ellis2012,Dixon1996} and references therein. 
The flat space analogue of the universal functional traces are the one-loop vacuum tensor integrals, which arise when the propagators in the integrand of a Feynman integral are expanded around vanishing external momenta. 
The resulting the logarithmically divergent tensor integrals can be evaluated in a closed form but require a subsequent contraction of the external momenta with the totally symmetrized product of metric tensors.

Despite the simplicity of this algorithm, in cases where the rank of the vacuum tensor integrals becomes high, the explicit tensor contraction leads to a fast proliferation of terms, rendering a brute-force implementation inefficient -- even for high performance computer algebra programs.
The high rank of the vacuum tensor integrals arise due to a high number of loop momenta in the numerator of a given Feynman integral that can have various origins. High powers of loop momenta in the numerator naturally arise in theories with a high number of derivatives in the interaction vertices. They also arise from massive vector field propagators as well as gauge field propagators in a general relativistic gauge. Finally, the propagator expansion around vanishing external momentum in addition increases the number of loop momenta in the numerator. 

The extraction of the one-loop divergences has been automatized by many computer algebra programs in different ways, using analytic as well as numerical methods \cite{Kueblbeck1990,Mertig1991,Hahn1999,Hahn2001,Denner2006,Binoth2009,Cullen2012,Mastrolia2012,Deurzen2014,Denner2017}.
Nevertheless, we believe that the combinatorial aspect presented in this note could improve the efficiency of these algorithms as it completely avoids to perform any tensor contraction.
While this problem practically mostly becomes relevant for higher derivative theories, the efficient combinatorial solution presented in this note might be integrated as a standard routine in existing computer algebra programs.

This note is structured as follows: In Sec. \ref{OneLoopInt}, we formulate the problem and introduce our notation.
In Sec. \ref{Combinatorics}, we provide a closed form combinatorial solution to the problem.
In Sec. \ref{Extensions} and Sec. \ref{ExtMassesLoops} we discuss the extension of the combinatorial algorithm to non-zero spin and to the case of multiple propagators with different masses, respectively.
Finally, in Sec. \ref{Conclusions}, we summarize our main results and comment on further applications. In \ref{DerCombinatorics} we provide more details on the derivation of combinatorial coefficients that enter the algorithm and in \ref{Example} we illustrate the general method by a concrete example of a higher derivative scalar field theory.

\section{One-loop integrals}
\label{OneLoopInt}
We work in flat spacetime with metric $\eta_{\mu\nu}=\mathrm{diag}(+,-,-,-)$ and consider a generic one-loop diagram with $n$ external legs.
\begin{figure}[h!]
\begin{center}
\begin{tikzpicture}
\begin{scope}[decoration={
	markings,
	mark=at position 0.5 with {\arrow{<}}}] 
\draw[black] (0,0) circle (0.5cm);
\draw[black,->] (0.2,0) arc (0:300:0.2);
\filldraw[black] (72:0.5) circle (0.7mm);
\draw[black,postaction={decorate}] (72:0.5) -- (72:1.3);
\draw[thick,dotted] (1,0.5) arc (15:35:1.5);
\node[black] at (72:1.5) {$k_{n-1}$};
\node[black] at (108:1.2) {$\ell+q_{n-1}$};
\filldraw[black] (144:0.5) circle (0.7mm);
\draw[black,postaction={decorate}](144:0.5) -- (144:1.3);
\node[black] at (144:1.6) {$k_n$};
\node[black] at (180:1) {$\ell$};
\filldraw[black] (216:0.5) circle (0.7mm);
\draw[black,postaction={decorate}] (216:0.5) -- (216:1.3);
\node[black] at (216:1.5) {$k_1$};
\node[black] at (252:0.8) {$\ell+q_1$};
\filldraw[black] (288:0.5) circle (0.7mm);
\draw[black,postaction={decorate}] (288:0.5) -- (288:1.3);
\node[black] at (288:1.6) {$k_2$};
\filldraw[black] (360:0.5) circle (0.7mm);
\node[black] at (330:1) {$\ell+q_2$};
\draw[black,postaction={decorate}](360:0.5) -- (360:1.3);
\node[black] at (360:1.6) {$k_3$};
\end{scope}
\end{tikzpicture}
\caption{Generic one-loop diagram with $n$ external legs and $n$ propagators. The arrows indicate the direction of the momentum flow: all external momenta $k_{1},\ldots,k_{n}$ are all incoming.}
\end{center}
\end{figure}
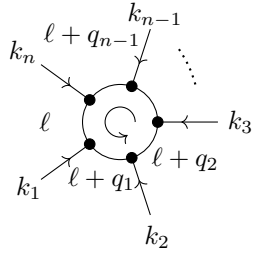

\noindent The inverses of the propagators (suppressing the $i\epsilon$),
\begin{align}
D_i:=\left(\ell+q_i\right)^2-m_i^2, \quad i=1,\ldots,n
\end{align}
are labeled by the combination of external momenta $q_i$ flowing in the $D_i$,
\begin{align}
q_i^{\mu}:=\sum_{j=1}^{i}k_j^{\mu},\qquad q_{n}^{\mu}=q_0^{\mu}=\sum_{j=1}^{n}k_j^{\mu}=0.\label{momq}
\end{align} 
Here, the $m_i$ denote the internal masses.
A generic scalar one-loop Feynman diagram corresponds to a sum of products of the form \footnote{We suppress the dependence on the masses in the arguments. Integrals for external particles with non-zero spin are discussed in Sec. \ref{Extensions}.}
\begin{align}
\tilde{I}:=&K\left(Q,M\right)I\left(Q,M\right),\label{IntFeyn}
\end{align}
where we have introduced the abbreviated notation $Q:=\{q_i\}$, $M=\{m_i\}$.
Here, $K\left(Q,M\right)$ is a sum of Lorentz scalars build from a product of kinematic invariants, where each term in the sum has the general structure
\begin{align}
K(Q,M):=c\prod_{1\leq i\leq j\leq  n}\left(q_i\cdot q_j\right)^{\rho_{ij}}m_{i}^{\mu_i},
\end{align}
with integer exponents $\rho_{ij}$ and $\mu_i$ and numerical constant $c$.
The second factor, $I(Q,M)$, is the actual loop integral
\begin{align}
I(Q,M):=\int\frac{\mathrm{d}^d\ell}{(2\pi)^{d}}\frac{N\left(Q,\ell\right)}{D\left(Q,M;\ell\right)}=\int\frac{\mathrm{d}^d\ell}{(2\pi)^{d}}(\ell^2)^\lambda\prod_{i=1}^{n}\frac{\left(q_i\cdot \ell\right)^{\sigma_i}}{D_i^{\delta_i}}\label{IntD},
\end{align}
with the numerator
\begin{align}
N\left(Q;\ell\right)=\left(\ell^2\right)^\lambda\prod_{i=1}^{n}\left(q_i\cdot \ell\right)^{\sigma_i},
\end{align}
and the denominator
 \begin{align}
D\left(Q,M;\ell\right)=\prod_{i=1}^{n}D_i^{\delta_i},
 \end{align}
with integer exponents $\lambda$, $\sigma_i$, $\delta_i$.
In the $\overline{\text{MS}}$ scheme, the calculation of counterterms only requires knowledge about the ultraviolet divergent part of the integrals \eqref{IntD}. 
Evaluating these integrals in $d=4-2\varepsilon$ dimensions, the one-loop divergences are isolated as poles in dimension $1/\varepsilon$ in the limit $\varepsilon\to0$.\footnote{In case of theories with massless particles an infrared (IR) regulating mass term $m_{\mathrm{IR}}^2$ might be introduced in the propagators.
The case of Feynman integrals with propagators involving different masses is discussed in Sec. \ref{ExtMassesLoops}.}
For the extraction of these pole terms, the calculation can be essentially simplified by expanding the propagators around vanishing external momenta $q_i=0$.
The expansion of a propagator $1/D_i$ including terms up to $\mathcal{O}\left(\ell^{-n}\right)$ can be compactly written as
\begin{align}
\frac{1}{D_i}=\sum_{0\leq 2\alpha+\beta+2\leq n }\left(
\begin{array}{c}
\alpha+\beta\\\alpha
\end{array}
\right)\frac{\left(-q_i^2\right)^\alpha\left(-2\ell\cdot q_i\right)^\beta}{\left(\ell^2-m_i^2\right)^{1+\alpha+\beta}}.\label{PropEx}
\end{align}
Since dimensional regularization annihilates all power-law divergences, we focus on the logarithmically divergent part.
By a simultaneous rescaling of the loop momenta $\ell\to\Lambda\ell$ and a subsequent counting of powers of $\Lambda$, we can infer the superficial degree of divergence $\chi_{\mathrm{div}}$.
Divergent integrals have $\chi_{\mathrm{div}}\geq0$. Logarithmically divergent integrals have $\chi_{\mathrm{div}}=0$ and integrals with $\chi_{\mathrm{div}}<0$ are finite.
Counting of the leading power of loop momenta in the numerator of \eqref{IntD} provides the required order up to which the propagators have to be expanded.

For the case of a single massive scalar field, which we are considering here, we have $m_i\equiv m$, $ i=1,\ldots,n$.
The cases of non-zero spin and different masses $m_i$ are discussed in Sec. \ref{Extensions} and Sec. \ref{ExtMasses}, respectively. 
After expansion of all propagators, the integral \eqref{IntD} is written as a sum of scalar vacuum integrals of the form
\begin{align}
I^{\mathrm{vac}}
=&\int\frac{\mathrm{d}^d\ell}{(2\pi)^{d}}\frac{(\ell^2)^\lambda\,\prod_{i=1}^{n}\left(q_i\cdot \ell\right)^{\sigma_i}}{\left(\ell^2-m^2\right)^\beta},\qquad 2\omega:=\sum_{i=1}^{n}\sigma_i.\label{LogDivInt}
\end{align}
Since the divergent integrals \eqref{LogDivInt} with odd powers of the loop momentum in the numerator vanish, we only consider integrals with even powers $2(\lambda+\omega)$ of $\ell$ in \eqref{LogDivInt}.
Among the integrals \eqref{LogDivInt}, the logarithmically divergent ones are extracted by power counting, leading to the constraint 
\begin{align}
\chi_{\mathrm{div}}=4+2(\omega+\lambda-\beta)=0.\label{logdivconst}
\end{align}
In the standard approach, the scalar products $(q_i\cdot \ell)$ in the integrals \eqref{LogDivInt} are broken up and the numerator is written in the form of a tensor contraction\footnote{The efficiency of the algorithm might be improved by first expressing all invariants $(qi1\cdot \ell)$ in terms of propagators $D_i$, which is always possible at the one-loop level.}
\begin{align}
N=(\ell^2)^\lambda\prod_{i=1}^{n}\left(q_i\cdot \ell\right)^{\sigma_i}=(\ell^2)^\lambda Q^{\mu_1\ldots\mu_{2\omega}}\ell_{\mu_1}\cdots\ell_{\mu_{2\omega}}.\label{QTensor}
\end{align}
The kinematic tensor $Q^{\mu_{1}\ldots\mu_{2\omega}}$ has even rank $2\omega$ and is constructed in terms of powers of different external momenta,
\begin{align}
Q^{\mu_1\ldots\mu_{2\omega}}=\underbrace{\left(q_1^{\mu_1}\cdots q_{1}^{\mu_{\sigma_1}}\right)}_{\sigma_1\text{ - times}}\underbrace{(q_{2}^{\mu_{\sigma_1+1}}\cdots q_{2}^{\mu_{\sigma_1+\sigma_2}})}_{\sigma_2 \text{- times}}\cdots\underbrace{(q_n^{\mu_{\sigma_1+\ldots+\sigma_{n-1}+1}}\cdots q_n^{\mu_{2\omega}})}_{\sigma_n \text{- times}}.\label{QKinTensor}
\end{align}
The remaining tensor vacuum integrals can be evaluated in a closed form \cite{Davydychev1991}, see also eq. (10.5) in \cite{Smirnov2012},
\begin{align}
I_{\mu_1\ldots\mu_{2\omega}}^{\mathrm{vac}}
={}&\int\frac{\mathrm{d}^d\ell}{\left(2\pi\right)^{d}}\frac{(\ell^2)^\lambda\,\ell_{\mu_1}\ldots\ell_{\mu_{2\omega}}}{\left(\ell^2-m^2\right)^\beta}\nonumber\\
={}&\frac{i(-1)^{\lambda+\omega+\beta}}{2^d\pi^{d/2}(m^2)^{\beta-\lambda-\omega-d/2}}\left[\eta_{\mathrm{sym}}^{\omega}\right]_{\mu_1\ldots\mu_{2\omega}}\frac{\Gamma(\lambda+\omega+d/2)\Gamma(\beta-\lambda-\omega-d/2)}{2^\omega\Gamma(\omega+d/2)\Gamma(\beta)}.\label{tenvac}
\end{align}
Here, $\left[\eta_{\mathrm{sym}}^{\omega}\right]_{\mu_1\ldots\mu_{2\omega}}$ is the totally symmetric product of $\omega$ metric tensors $\eta_{\mu\nu}$,
\begin{align}
\left[\eta_{\mathrm{sym}}^{\omega}\right]_{\mu_1\ldots\mu_{2\omega}}:=(2\omega-1)!!\,\eta_{(\mu_1\mu_2}\cdots\eta_{\mu_{2\omega-1}\mu_{2\omega})},\label{defgsym}
\end{align}
where the round brackets denote total symmetrization with unit weight among the $2\omega$ enclosed indices
and the overall factor of ${(2\omega-1)!!=(2\omega)!/(2^\omega \omega!)}$ ensures that each of the ${(2\omega-1)!!}$ terms in the sum has unit coefficient.

In particular, the constraint \eqref{logdivconst} implies that logarithmically divergent  integrals \eqref{tenvac} in ${d=4}$ spacetime dimensions correspond to 
\begin{align}
\left.I_{\mu_1\ldots\mu_{2\omega}}^{\mathrm{vac}}\right|^{\mathrm{div}}=\frac{i}{(4\pi)^2\varepsilon}\frac{1}{P_{\omega}(4)}.\label{LogDivInt2}
\end{align}
Here, $P_{\omega}(4)$ is the $d=4$ case of the polynomial in spacetime dimension
\begin{align}
P_\omega(d)=\prod_{i=1}^{\omega}\left[d+2(i-1)\right]=\frac{2^{\omega}\Gamma(\omega+d/2)}{\Gamma(d/2)},\label{Pold}
\end{align}
which is defined as the trace of $\left[\eta_{\mathrm{sym}}^{\omega}\right]_{\mu_1\ldots\mu_{2\omega}}$, \footnote{A more detailed derivation of the factors $(2\omega-1)!!$ in \eqref{defgsym} and $P_{\omega}(d)$ in \eqref{Pold} is provided in \ref{DerCombinatorics}.},
\begin{align}
P_\omega(d):=\left[\eta_{\mathrm{sym}}^{\omega}\right]_{\mu_1\ldots\mu_{2\omega}}\eta^{\mu_1\mu_2}\ldots\eta^{\mu_{2\omega-1}\mu_{2\omega}}.\label{NormCond}
\end{align}  
The last step in the calculation is to perform the tensor contraction 
\begin{align}
Q^{\mu_1\ldots\mu_{2\omega}}\left[\eta_{\mathrm{sym}}^{\omega}\right]_{\mu_1\ldots\mu_{2\omega}}.\label{Qgsym}
\end{align}
This task is usually performed by symbolical computer algebra programs, such as e.g.  \texttt{FORM} \cite{Vermaseren2000}, \texttt{Cadabra} \cite{Peeters2007} or the \texttt{Mathematica} tensor algebra package \texttt{xTensor} \cite{MartinGarcia}.
In particular, \texttt{FORM} can handle a large number of contractions with a high performance.
Nevertheless, for higher derivative theories, the high number of derivatives in the vertices in general lead to a high number of loop momenta in the numerators of Feynman integrals.
Moreover, additional loop momenta arise in the numerator due to the propagator expansion \eqref{PropEx}.
Ultimately, this might lead to very high-rank tensors $Q^{\mu_1\ldots\mu_{2\omega}}$ which render this brute-force implementation inefficient. 

\section{Combinatorics of Feynman integrals}
\label{Combinatorics}
The first observation is that the introduction of the tensors $Q^{\mu_1\ldots\mu_{2\nu}}$ and $\left[\eta_{\mathrm{sym}}^{\nu}\right]_{\mu_1\ldots\mu_{2\nu}}$ as an intermediate step of the calculation can be completely avoided.
The result of the contraction \eqref{Qgsym} is a sum of products of powers of invariants,
\begin{align}
Q^{\mu_1\ldots\mu_{2\nu}}\left[\eta_{\mathrm{sym}}^{\nu}\right]_{\mu_1\ldots\mu_{2\nu}}=\sum_{k\in\mathcal{P}(2\omega:\sigma_{ij})}C_{k}\prod_{1\leq i\leq j\leq n}\left(q_i\cdot q_j\right)^{\sigma_{ij}^{k}}.\label{tencont}
\end{align}
The sum over $k$ in \eqref{tencont} runs over all partitions ${\mathcal{P}(2\omega:\sigma_{ij})}$ of $2\omega=\sum_{i}\sigma_i$ into the $n(n+1)/2$ non-negative integers $\sigma_{ij}=\sigma_{ji}$ under the $n$ constraint equations $\sigma_i=2\sigma_{ii}+\Sigma_i$ where ${\Sigma_i:=\sum_{j\neq i=1}^n\sigma_{ij}}$.
Each partition $k\in {\mathcal{P}(2\omega:\sigma_{ij})}$ with fixed numbers $\sigma_{ij}^k$ comes with a numerical coefficient $C_k$ that needs to be determined. 
This is a purely combinatorial problem, which can be solved in a closed form. At a fundamental level, this problem is related to the representation theory of the symmetric group, but we derived in a more elementary way by counting the number of possible pairings among the external momenta. 
The solution to the problem can be subdivided into three steps:
\begin{enumerate}
	\item Find all integer partitions ${\mathcal{P}(2\omega:\sigma_{ij})}$ of ${2\omega=\sum_{i=1}^n\sigma_i}$ into the ${n(n+1)/2}$ non-zero integers ${\sigma_{ij}=\sigma_{ji}}$ compatible with the $n$ constraint equations ${\sigma_i=2\sigma_{ii}+\Sigma_i}$, given that the non-negative integers $\sigma_i\geq0$ are known.
	The desired partitions are given by all non-negative integer solutions of the system of $n$ linear equations
	\begin{align}
	\sigma_{1}=&2\sigma_{11}+\sigma_{12}+\ldots +\sigma_{1n},\nonumber\\
	\sigma_{2}=&\sigma_{21}+2\sigma_{22}+\ldots+\sigma_{2n},\nonumber\\
	&\vdots\nonumber\\
	\sigma_n=&\sigma_{n1}+\sigma_{n2}+\ldots+2\sigma_{nn}.\label{SysPart}
	\end{align}
	Each solution is a set of $n(n+1)/2$ integers $\sigma_{ij}^k$ associated with the partition $k$ and therefore generates a different combination of invariants $\prod_{1=i<j=n}(q_{i}\cdot q_{j})^{\sigma_{ij}^{k}}$.
	Finding all partitions is easily accomplished by a computer program. The program reads in the $n$ numbers $\sigma_i$ and returns a list of $k$ elements, each element corresponding to an ordered list of $n(n+1)/2$ numbers $\sigma_{ij}^k$.\\
	\item 
	Having found all partitions ${\mathcal{P}(2\omega:\sigma_{ij})}$, for each partition $k\in {\mathcal{P}(2\omega:\sigma_{ij})}$, we must find the coefficient $C_{k}$ as a function of the $\sigma_{ij}^k$.
	The $C_{k}$ is a purely combinatorial factor as it amounts of counting the number of ways the $2\omega$ external momenta $q_i$, $i=1,\ldots,n$ can be paired among each other to form the invariants ${\prod_{1\leq i\leq j\leq n}(q_i\cdot q_j)^{\sigma_{ij}^k}}$. 
	The coefficient $C_k$ as a function of the $\sigma_{ij}$ reads
	\begin{align}
	C_k=\frac{\prod_{j=1}^{n}\left[\Sigma_j!
		\left(
		\begin{array}{c}
		2\sigma_{jj}+\Sigma_j
		\\
		\Sigma_j
		\end{array}
		\right)
		(2\sigma_{jj}-1)!!\right]}{\prod_{1\leq i< j\leq n}\sigma_{ij}!}.\label{Ck}
	\end{align}
A derivation of \eqref{Ck} is provided in \ref{DerCombinatorics}.
	\item Summing over all partitions $k\in {\mathcal{P}(2\omega:\sigma_{ij})}$ and taking into account the contribution from the dimensional polynomial \eqref{Pold}, we obtain for the logarithmically divergent integral \eqref{LogDivInt} with ${d+2(\lambda+\omega-2\beta)=0}$ in $d=4$,
	\begin{align}
	\left.I^{\mathrm{vac}}\right|^{\mathrm{div}}=\frac{i}{\varepsilon(4\pi)^2}\sum_{k\in \mathcal{P}(2\omega:\sigma_{ij})}\frac{C_k}{P_{\omega}(4)}\prod_{1\leq i\leq j\leq n}(q_i\cdot q_j)^{\sigma_{ij}^{k}}.\label{FinRes}
	\end{align}
\end{enumerate}

\noindent The combinatorial approach presented in this paper can be easily implemented on a computer: First a power counting function is applied to the numerator of a given Feynman integral of the form \eqref{IntD}, which determines the required order to which the propagators have to be expanded.
After the propagator expansion, the power counting function is applied again to the integrands of the resulting sum of integrals (which are all of the form \eqref{LogDivInt}) in order to select the logarithmically divergent ones.
For each of these logarithmically divergent integrals, a function reads off the exponents $\sigma_{i}$ of the $n$ scalar products $(q_i\cdot \ell)^{\sigma_i}$.
Next, all integer partitions of the sum $2\omega=\sum_{i=1}^{n}\sigma_i$ into the $n(n+1)/2$ exponents $\sigma_{ij}$ of the $\omega$ invariants $(q_i\cdot q_{j})$ are generated by solving the system \eqref{SysPart}.
Then, for each partition, the corresponding combinatorial coefficient \eqref{Ck} is computed and the pole in dimension of the simple vacuum one-loop integral is extracted by \eqref{LogDivInt2}.
Finally, summing over all partitions gives the desired result \eqref{FinRes}.
Doing this for all logarithmically divergent integrals obtained from the propagator expansion and summing over all contributions gives the logarithmically divergent part of the original Feynman integral \eqref{IntFeyn}. An explicit illustrative example of this procedure is provided in \ref{Example}.

\section{Extensions of the algorithm to nonzero spin}
\label{Extensions}
The algorithm in Sec. \ref{OneLoopInt} and Sec. \ref{Combinatorics} was presented for a single scalar field.
Its extension including spin one-half, spin-one, and spin-two particles (in fact for arbitrary spin) is possible without great modifications.
Since the combinatorial algorithm relies on counting the number of different pairings between vectors, for Feynman diagrams with non-scalar external legs we need to introduce a set of auxiliary vectors, which absorb possible Lorentz and/or spinor indices in the corresponding Feynman integrals. 
In the following discussion, we work in $d=4$ spacetime dimensions and make explicit use of dimensional dependent identities (DDI) only valid in $d=4$.
This is legitimate as long as we are dealing only with the calculation of one-loop ultraviolet divergences which are proportional to $1/\varepsilon$ using dimensional regularization in $d=4-2\varepsilon$ dimensions, but must be generalized for higher loops and finite contributions in which case $\mathcal{O}\left(\varepsilon\right)$ corrections of the DDI's become relevant.

\subsection{Spin $1/2$}
\label{spinonehalf}
Extending the algorithm to spin-$1/2$ particles, one might encounter Feynman integrals with numerators that involve Lorentz contractions with and among bilinear ``spinor tensor chains'' such as e.g. 
\begin{align}
\cdots\left(\bar{u}_{i}U^{\mu_1\ldots\mu_n}v_{j}\right)\cdots\left(\bar{w}_kV_{\mu_1\cdots\mu_n}z_{l}\right)\ldots.
\end{align}
In the above expression, we have suppressed spinor indices of the Dirac spinors $u_{i}$, $\bar{v}_{i}$, $w_{i}$, $\bar{z}_{i}$, etc., and have denoted adjoint spinors with an overline. We denote a bilinear spinor tensor chain in which all spinor indices are fully contracted by an enclosing angular bracket. 
In general, any external four component Dirac spinor $u^{a}_{s}(k_i)$ is characterized by its associated momentum $k_{i}$, its spinor index $a=1,\ldots,4$ and its polarization index $s=1,2$.
In the following discussion, we suppress the polarization index $s$. Whenever we suppress in addition the spinor index $a$, we indicate the dependence on the momentum $k_{i}^{\mu}$ by a corresponding subindex, i.e. $\bar{u}_{i}:=\bar{u}(k_{i})$.

Each spinor chain involves a Dirac matrix-valued tensor $\tensor{\left[U^{\mu_1\ldots\mu_n}\right]}{_{a}^{b}}$, which acts on the space of Dirac spinors and contains powers of the Dirac matrices $\tensor{\left[\gamma^{\mu}\right]}{_{a}^{b}}$, $\tensor{[\gamma^{5}]}{_{a}^{b}}$, powers of the external momenta $k_{i}^{\mu}$ and powers of the loop momentum $\ell^{\mu}$.
The Dirac matrices satisfy the Clifford algebra
\begin{align}
\{\gamma^{\mu},\gamma^{\nu}\}=2\eta^{\mu\nu}\mathbf{1},\qquad \{\gamma^{\mu},\gamma^5\}=0,\label{CA}
\end{align}
where $\mathbf{1}=[\delta]^{a}_{b}$ denotes the unit matrix in spinor space. In $d=4$ spacetime dimensions, a possible representation of $\gamma^{5}$ is given in terms of the totally antisymmetric Levi-Civita tensor (defined such that $\varepsilon_{0123}=1$),
\begin{align}
\gamma^{5}=\frac{i}{4!}\epsilon_{\mu\nu\rho\sigma}\gamma^{\mu}\gamma^{\nu}\gamma^{\rho}\gamma^{\sigma}.
\end{align}  
Furthermore, in $d=4$, a basis in the space of Dirac bilinears can be compactly written in terms of the five Dirac matrix-valued tensors $\Gamma^{X}$, $X=S,V,T,P,A$, which are explicitly defined by
\begin{align}
\Gamma^{S}:=\mathbf{1},\quad\Gamma^{V}:=\gamma^{\mu},\quad \Gamma^{T}:=\sigma^{\mu\nu},\quad\Gamma^{P}:=\gamma^{5},\quad\Gamma^{A}:=\gamma^{5}\gamma^{\mu},\label{DiracBasis}
\end{align} 
Here, the rank-two antisymmetric Dirac tensor $\sigma^{\mu\nu}$ is defined by
\begin{align}
\sigma^{\mu\nu}:=i\gamma^{[\mu}\gamma^{\nu]}.
\end{align}
The indices $S$, $V$, $T$, $P$, $A$, which label the elementary Dirac matrix-valued tensors \eqref{DiracBasis}, stand for ``scalar'', ``vector'', ``tensor'', ``pseudoscalar'', and ``axial vector'' according to their transformation properties under Lorentz transformations
\begin{align}
S_{(ij)}:=\left(\bar{u}_{i}\Gamma^{S}u_{j}\right)\to\left(\bar{u}_{i}\Gamma^{S}u_{j}\right)'={}&\left(\bar{u}_{i}\Gamma^{S}u_{j}\right)\label{EDS},\\
V_{(ij)}^{\mu}:=\left(\bar{u}_{i}\Gamma^{V}u_{j}\right)\to\left(\bar{u}_{i}\Gamma^{V}u_{j}\right)'={}&\Lambda\left(\bar{u}_{i}\Gamma^{V}u_{j}\right),\\
T_{(ij)}^{\mu\nu}:=\left(\bar{u}_{i}\Gamma^{T}u_{j}\right)\to\left(\bar{u}_{i}\Gamma^{T}u_{j}\right)'={}&\Lambda\left(\bar{u}_{i}\Gamma^{T}u_{j}\right)\Lambda,\\
P_{(ij)}:=\left(\bar{u}_{i}\Gamma^{P}u_{j}\right)\to\left(\bar{u}_{i}\Gamma^{P}u_{j}\right)'={}&\textrm{det}(\Lambda)\left(\bar{u}_{i}\Gamma^{P}u_{j}\right),\\
A_{(ij)}^{\mu}:=\left(\bar{u}_{i}\Gamma^{A}u_{j}\right)\to\left(\bar{u}_{i}\Gamma^{A}u_{j}\right)'={}&\textrm{det}(\Lambda)\Lambda\left(\bar{u}_{i}\Gamma^{A}u_{j}\right),\label{EDC}
\end{align}
where we have suppressed the Lorentz indices of the Lorentz transformation matrices $\tensor{\Lambda}{^{\mu}_{\nu}}$.
Therefore, in $d=4$, each bilinear Dirac chain can be reduced to a linear combination of the elementary bilinear Dirac chains \eqref{EDS}-\eqref{EDC} involving only the basis elements \eqref{DiracBasis}.
The explicit reduction of the individual bilinear Dirac chains can, e.g., be performed iteratively by repeated use of $(\gamma^{5})^2=\mathbf{1}$ and the Chisholm identities, which reduce chains of three of more Dirac matrices to sums of the elementary basis chains in \eqref{DiracBasis},
\begin{align}
\gamma^{\mu}\gamma^{\nu}\gamma^{\rho}={}&\eta^{\mu\nu}\gamma^{\rho}+\eta_{\nu\rho}\gamma^{\mu}-\eta^{\rho\mu}\gamma^{\nu}-i\tensor{\varepsilon}{^{\mu\nu\rho}_{\sigma}}\gamma^{\sigma}\gamma^{5},\\
\gamma^{\mu}\gamma^{\nu}\gamma^{5}={}&\eta^{\mu\nu}\gamma^{5}-\frac{1}{2}\tensor{\varepsilon}{^{\mu\nu}_{\rho\sigma}}\sigma^{\rho\sigma},
\end{align}
and contraction identities among $\gamma$-matrices, which follow directly from the Clifford algebra \eqref{CA}. In addition, the Fierz identities might be used for to swap the momentum dependence of the external spinors in products of Dirac bilinears, see e.g. \cite{Pal2007} for more details.

After the reduction of all individual bilinear spinor chains, there remain the Lorentz contractions of elementary bilinear Dirac chains $V_{(ij)}^{\mu}$, $A_{(ij)}^{\mu}$ and $T_{(ij)}^{\mu\nu}$ either with external momenta $q_{i}^{\mu}$, the loop momentum $\ell^{\mu}$, or with other elementary bilinear Dirac chains. Ultimately, all numerators in a given Feynman integral can therefore be reduced to a product of Lorentz scalars by introducing a number of auxiliary vectors of the form $V_{(ij)}^{\mu}$, $A_{(ij)}^{\mu}$ and possible `tensor contraction chains', such as 

\begin{align}
T_{(ijk)}^{\mu}:={}&T^{\mu}_{(ij)\rho}q^{\rho}_{k},\\
TT_{(ijklm)}^{\mu}:={}&T^{\mu}_{(ij)\rho}T^{\rho}_{(kl)\nu}q^{\nu}_{m},\\
TTT_{(ijklmns)}^{\mu}:={}&T^{\mu}_{(ij)\rho}T^{\rho}_{(kl)\nu}T^{\nu}_{(mn)\lambda}q^{\lambda}_{s},\\
\vdots\,\;&\nonumber
\end{align}
Whenever these auxiliary vectors appear in scalar products with the loop momentum $\ell^{\mu}$ in the numerator, they have to be considered as additional external vectors in the combinatorial algorithm. It is clear that the number of required auxiliary vectors grows with the number of external legs in general and with the number of external spinor states in particular.

\subsection{Spin $1$}
\label{spinone}
Incorporating spin-$1$ vector fields into the algorithm is straightforward. In this case, the numerators of the Feynman integrals involve polarization vectors for incoming external vector particles
\begin{align}
\varepsilon_{i}^{\mu}:=\varepsilon^{\mu}(k_i):=\varepsilon^{\mu}_{\lambda}(k_i),
\end{align}
and their complex conjugates $\varepsilon_{i}^{*\mu}$ for outgoing external vector particles.
In our notation, we suppress the little-group index $\lambda$, which labels the physical polarizations and runs from $\lambda=0,+,-$ for massive vector fields and from $\lambda=+,-$ for massless vector fields. From the viewpoint of the combinatorial off-shell algorithm, the $\varepsilon^{\mu}(k_i)$ and their conjugates $\varepsilon^{*}_{\mu}(k_i)$ are treated as separate additional external vectors on equal footing with the external momenta $k_{i}^{\mu}$. 
Thus, in a $n$-point diagram, there are at most $n$ different additional external vectors to be paired.\\

\subsection{Spin $2$}
\label{spintwo}
The extension of the combinatorial algorithm to spin-$2$ fields can essentially be reduced to the spin-$1$ case. The symmetric traceless spin-$2$ polarization tensors are defined as
\begin{align}
\epsilon^{\mu\nu}_{\lambda}(k_i)=\epsilon^{\mu\nu}_{i},
\end{align}
where we have again suppressed the polarization index $\lambda$ ($\lambda=+,\times$ for the massless case and $\lambda=+,\times,1,2,3$ for the massive case) and indicated the dependence on the external momentum $q_i$ by a subindex $i$. 
Since these symmetric tensors $\epsilon^{\mu\nu}_{i}$ and their complex conjugates $\epsilon^{*\mu\nu}_{i}$ only appear once in a given Feynman integral, without loss of generality, we can write them as direct product of two vectors $\epsilon^{\mu}_{i}$ and $\tilde{\epsilon}^{\mu}_{i}$,
\begin{align}
\epsilon_{i}^{\mu\nu}=\epsilon^{\mu}_{i}\tilde{\epsilon}^{\nu}_{i}.\label{DecEps}
\end{align} 
In view of \eqref{DecEps}, contractions $\epsilon^{\mu\nu}_iq_{j\mu}\ell_{\nu}$ and $\epsilon^{\mu\nu}_i\ell_{\mu}\ell_{\nu}$ involving the loop momentum $\ell$, lead to scalar products of the form 
\begin{align}
\epsilon^{\mu\nu}_ik_{j\mu}\ell_{\nu}=(\epsilon_i\cdot k_j)(\tilde{\epsilon}_i\cdot \ell),\qquad \epsilon^{\mu\nu}_i\ell_{\mu}\ell_{\nu}=(\epsilon_i\cdot \ell)(\tilde{\epsilon}_i\cdot \ell).\label{SPT}
\end{align}
From the viewpoint of the combinatorial algorithm the $\epsilon_i^{\mu}$, $\tilde{\epsilon}_i^{\mu}$, $\epsilon^{*\mu}_{i}$ and $\tilde{\epsilon}^{*\mu}_{i}$ are just considered as different additional external vectors. Loop integration turns scalar products such as \eqref{SPT} into invariants such as e.g.
\begin{align}
(\epsilon_i\cdot k_j)(\tilde{\epsilon}_i\cdot k_k).
\end{align}
Since the $\epsilon_{i}$, $\tilde{\epsilon}_i$ are labeled uniquely, the contractions of external momenta with the original polarization tensors $\epsilon^{\mu\nu}_{i}$ can be reconstructed from combining those two scalar products which involve the vectors $\epsilon_{i}^{\mu}$ and $\tilde{\epsilon}_{i}^{\mu}$ which share the same index $i$, e.g.
\begin{align}
(\epsilon_i\cdot k_j)(\tilde{\epsilon}_i\cdot k_k)=\epsilon_{i}^{\mu\nu}k_{j\mu}k_{k\nu}.
\end{align} 
In this way the dependence of the integral on the original polarization tensors $\epsilon_{i}^{\mu\nu}$ can be unambiguously reconstructed. 
The procedure outlined in Sec.~\ref{spinonehalf} - Sec.~\ref{spintwo} can be extended to higher spin fields. In general, for the combinatorial algorithm, the auxiliary vectors do not have to be constructed from the physical polarization tensors but can be arbitrary external vectors, which just act as distinguishable `placeholders' introduced to absorb all Lorentz and spinor indices.  

\section{Extension to multiple propagators and higher loops}
\label{ExtMassesLoops}
In this section, we briefly discuss the extension of the combinatorial algorithm to diagrams involving propagators with different masses and the extension to higher loops.

\subsection{Extension to integrals with multiple different propagators}
\label{ExtMasses}
The extension of the algorithm to diagrams involving different propagators is not difficult. Different propagators differ in two aspects. First, propagators with different spin have a different index structure which, however, only enters the numerator of the Feynman integral and can be incorporated in the combinatorial algorithm as described in Sec. \ref{Extensions}. Second, different propagators (of equal or different spin) might differ in their masses $m_i$. Performing the propagator expansion \eqref{PropEx} for each of these propagators, the denominator of the integrand of the resulting Feynman integrals acquires the schematic structure
\begin{align}
&\prod_{i=1}^{s}\frac{1}{(\ell^2+m_{i}^2)^{\kappa_i}}.
\end{align} 
Due to the product of propagators with different masses, we cannot directly make use of the general single-mass formula \eqref{tenvac} in order to perform the loop integral. However, we can first reduce a multi-mass integral to a single-mass integral and then make again use of \eqref{tenvac}. This reduction can be accomplished by introducing an infinitesimal IR regulating mass term in all propagators $m_i^2\to m_i^2+m_{\mathrm{IR}}^2$ and by expanding around zero masses $m_{i}=0$, $i=1,\ldots,s$. The expansion of the single $\kappa_i$-fold propagator $1/D_i^{\kappa_i}$ around vanishing mass $m_i=0$ up to $n$th order yields
\begin{align}
&\frac{1}{(l^2+m_{i}^2+m_{\mathrm{IR}}^2)^{\kappa_i}}=\frac{1}{(l^2+m_{\mathrm{IR}}^2)^{\kappa_i}}\left[\sum_{j=0}^{n}\left(\begin{array}{c}k_i+j-1\\j\end{array}\right)\frac{(-m_i^2)^j}{(\ell^2+m_{\mathrm{IR}}^2)^{j}}\right].\label{multmassexp}
\end{align} 
Performing the expansion \eqref{multmassexp} for all propagators $1/D_i^{\kappa_i}$, $i=1,\ldots,s$ and selecting the logarithmically divergent parts, the problem is effectively reduced to a single-mass integral for which \eqref{tenvac} can again be used with $m=m_{\mathrm{IR}}$.

\subsection{Extension to higher loops}
\label{ExtLoops}
The extension to higher loops is much more complicated for two reasons. First, the DDI used in $d=4$ for the one-loop calculation are no longer valid, as their deviations are of order $\mathcal{O}(\varepsilon)$. Consider for example a two-loop calculation. The leading pole from the UV divergent integrals is $1/\varepsilon^2$. Therefore $\mathcal{O}(\varepsilon)$ contributions from the expansion of the DDI can combine to give UV divergent $1/\varepsilon$ contributions. Second, staring from two-loops, in general  subdivergences appear which spoil the simple power counting used to truncate the propagator expansion and to extract the logarithmically divergent parts at the one-loop level. The extraction of UV divergences and renormalization at higher loops can be carried out in a systematic and recursive way by the $R$-operation \cite{Bogoliubov1957,Hepp1966,Zimmermann1968,Zimmermann1969,Collins1974}.

\section{Conclusions}
\label{Conclusions}
In this note we have presented a closed form combinatorial algorithm for the evaluation of one-loop divergences in Feynman diagrammatic calculations. The combinatorial approach is much more efficient and faster than a `brute force' implementation, as the unnecessary generation of a large number of intermediate tensors and their subsequent contractions is completely avoided. We presented the combinatorial algorithm for a single scalar field and discussed its extension to integrals for external particles with non-zero spin and to integrals involving multiple propagators with different masses. We also briefly commented on the extension to higher loops. We provided a derivation of the combinatorial formulas entering the algorithm, performed several checks of the combinatorial coefficient \eqref{Ck}, and illustrated the application of the algorithm by a concrete example of a higher derivative scalar field theory. The highest efficiency gain for the extraction of one-loop divergences within to the combinatorial approach presented in this note can be expected for higher derivative theories. Since high-rank vacuum tensor integrals also arise in a given truncation of an effective field theory or even in an ordinary second-order field theory, the combinatorial algorithm might also be implemented as a standard routine in existing computer algebra programs designed for the automated evaluation of Feynman diagrams.   

\section{Acknowledgements}
I thank Lance Dixon for discussions and Stefan Dittmaier and Michael Ruf for helpful comments on the first draft of this paper.   

\appendix
\section{Derivation of combinatorial formulas}
\label{DerCombinatorics}
In this section we provide the derivation of several combinatorial expressions arising in the algorithm for the calculation of the one-loop divergences.

\subsection{Factor of $(2\omega-1)!!$ in $[\eta_{\mathrm{sym}}^\omega]_{\mu_1\ldots\mu_{2\omega}}$}
\label{GsymFac}
The factor of $(2\omega-1)!!$ arising in the definition \eqref{defgsym} of $[\eta_{\mathrm{sym}}^\omega]_{\mu_{1}\ldots\mu_{2\omega}}$ corresponds to the number of ways to form $\omega$ orderless (since the metric tensor is symmetric ${\eta_{\mu\nu}=\eta_{\nu\mu}}$) pairs out of $2\omega$ indices. Having fixed one index to form the first pair, there remain $2\omega-1$ indices to pair with. Once the first pair has been formed, again an arbitrary index is fixed to form the second pair by combining with one of the remaining $2\omega-3$ indices, etc. Thus, the number of possible ways to pair the $2\omega$ indices is
\begin{align}
(2\omega-1)(2\omega-3)\cdots1=(2\omega-1)!!.
\end{align}

\subsection{Polynomial in spacetime dimension $P_{\omega}(d)$}
\label{DimPol}
The explicit form \eqref{Pold} of the polynomial $P_\omega(d)$ can be derived from  \eqref{NormCond}. For a given $\omega$, among the sum of products of metric tensors, there is only one combination of the product of $\omega$ metric tensors which is identical to the one we are contracting with. Therefore, there is only one combination where each of the $\omega$ metric tensors are traced, giving rise to the leading monomial $d^\omega$ of $P_\omega(d)$ with unit coefficient. For the next-to-leading monomial, there are $\omega-2$ metric tensors which are traced, while two metric tensors contract with crossed indices such as e.g. in $\eta^{\mu_1\mu_2}\eta^{\mu_3\mu_4}\eta_{\mu_1\mu_3}\eta_{\mu_2\mu_4}=d$. The polynomial $P_{\omega}(d)$ might therefore be constructed iteratively. Staring with $\omega=1$, there is only one contraction, leading to a power of $d$. Next, for $\omega=2$, there is only one way to contract all metric tensors to a trace resulting in the leading power of $d^2$ and two ways to contract with crossed indices, resulting in a contribution $2d$. This continues iteratively for each new metric tensor added, resulting in 
\begin{align}
P_\omega(d)={}&d(d+2)(d+4)\cdots(d+2(\omega-1))\\
={}&\prod_{i=1}^{\omega}\left[d+2(i-1)\right]=\frac{2^{\omega}\Gamma(\omega+d/2)}{\Gamma(d/2)}.\label{dimpol}
\end{align}

\subsection{Combinatorial weight factor $C_{k}$}
\label{CombWeight}
In order to derive the general formula \eqref{Ck} with $\Sigma_{i},\sigma_{ij}\in \mathbb{N}_0$, $i=1,\ldots n$, we investigate the individual factors in the product and recall the definition $\Sigma_{j}=\sum_{i\neq j=1}^n \sigma_{ij}$,
\begin{align}
C_k={}&\frac{\prod_{j=1}^{n}\left[\Sigma_j!
	\left(
	\begin{array}{c}
	2\sigma_{jj}+\Sigma_j
	\\
	\Sigma_j
	\end{array}
	\right)
	(2\sigma_{jj}-1)!!\right]}{\prod_{1\leq i< j\leq n}\sigma_{ij}!}={}\frac{\prod_{j=1}^{n}\left[2^{-\sigma_{jj}}\left(2\sigma_{jj}+\Sigma_{j}\right)!\right]}{\prod_{1\leq i\leq j\leq n}\sigma_{ij}!}
.\label{Ck2}
\end{align}
We count the number of ways we can pair $n$ different external vectors $q_{i}$,  ${i=1,\ldots,n}$, each with multiplicity $\sigma_{i}$ to form the $n(n+1)/2$ different scalar products $(q_{i}\cdot q_{j})$, each with multiplicity $\sigma_{ij}$, i.e. we have $n$ different sets $q_{i}$, each with $\sigma_{i}$ identical elements and want to know how many ways there are to form $n(n+1)/2$ different sets of pairs $(q_{i}\cdot q_{j})$, each with $\sigma_{ij}$ identical elements. 

\noindent We first explain the individual factors in \eqref{Ck2} and later apply the formula explicitly to several illustrative examples.
The $\Sigma_{j}$ count the number of $q_{j}$ which are paired with a different vector $q_{i}$, $i\neq j$. 
Therefore, the binomial factor 
\begin{align}
\left(\begin{array}{c}2\sigma_{jj}+\Sigma_{j}\\\Sigma_{j}\end{array}\right)=\frac{(2\sigma_{jj}+\Sigma_{j})!}{\Sigma_{j}!(2\sigma_{jj})!}=2^{-\sigma_{jj}}\frac{(2\sigma_{jj}+\Sigma_{j})!}{\Sigma_{j}!\sigma_{jj}!(2\sigma_{jj}-1)!!},
\end{align}
counts the number of ways we can choose the $\Sigma_{j}$ vectors $q_j$, which are paired with a different vector $q_{i}$, $i\neq j$ out of the total number $2\sigma_{jj}+\Sigma_{j}$ of vectors $q_{j}$. For each of these $\Sigma_{j}$ vectors $q_{j}$, there are $\Sigma_{j}!$ ways to pair with a different vector $q_{i}$ with $i\neq j$. In order not to overcount, we have to divide by the $\sigma_{ij}!$ with $i<j$.
The remaining $2\sigma_{jj}$ vectors $q_{j}$ must pair among themselves and there are $(2\sigma_{jj}-1)!!$ possible ways to do so.
In the following subsection, we illustrate how the combinatorial coefficient \eqref{Ck} can be tested for several explicit simple cases with increasing complexity.

\subsubsection{One external momentum}
\label{Apppart1}
We first look at the case with a single external momentum $q_1$. In this case the numerator of \eqref{LogDivInt} is just $N(q_1)=(q_1\cdot \ell)^{\sigma_1}$ with $2\omega=\sigma_1=12$ and the system to be solved in order to obtain all non-negative integer partitions is trivial
\begin{align}
12=2\sigma_{11}.
\end{align}
The only non-negative integer solution together with the combinatorial weight factor  are collected in Table \ref{OneExtLeg}
\begin{table}[h!]
	\begin{center}
	\begin{tabular}{c|c|c}
		$k$ & $(\sigma_{11})$&$C_k$\\
		\hline
		$1$ & $(6)$ & $10395$
	\end{tabular}
	\end{center}
\caption{All partitions $k$ together with the corresponding values of the $\sigma_{ij}$ and the combinatorial weight factor $C_k$ for $2\omega=\sigma_1=12$.}
\label{OneExtLeg}
\end{table}
\noindent As a check, we derive this number in the standard way by performing the tensor contraction \eqref{Qgsym},
\begin{align}
q_{1}^{\mu_1}q_{1}^{\mu_2}\ldots q_{1}^{\mu_{12}}[\eta_{\mathrm{sysm}}^{6}]_{\mu_1\mu_2\ldots \mu_{12}}
=&10395(q_1\cdot q_1)^6.
\end{align}
We read off the combinatorial coefficient $C_{1}=10395$, which is in agreement with the result obtained by the combinatorial formula.

\subsubsection{Two external momenta}
\label{Apppart2}
Next, we consider again the integral \eqref{LogDivInt} for a fixed number of loop momenta $2\omega=12$ with two external momenta $q_1$ and $q_2$ in the numerator $N(q_1,q_2)=(q_1\cdot \ell)^{\sigma_1}(q_2\cdot \ell)^{\sigma_2}$ with $2\omega=\sigma_1+\sigma_2=6+6$. The system to be solved in order to obtain all non-negative integer partitions is
\begin{align}
6=&2\sigma_{11}+\sigma_{12},\nonumber\\
6=&2\sigma_{22}+\sigma_{12}.
\end{align}
All four non-negative integer solution together with their combinatorial weight factors  are collected in Table \ref{TwoExtLeg}
\begin{table}[h!]
	\begin{center}
		\begin{tabular}{c|c|c}
			$k$ & $(\sigma_{11},\sigma_{12},\sigma_{22})$&$C_k$\\
			\hline
			$1$ & $(0,6,0)$ & $720$\\
			$2$ & $(1,4,1)$ & $5400$\\
			$3$ & $(2,2,2)$ & $4050$\\
			$4$ & $(3,0,3)$ & $225$
		\end{tabular}
	\end{center}
	\caption{All partitions $k$ together with the corresponding values of the $\sigma_{ij}$ and the associated combinatorial weight factor $C_k$ for $2\omega=\sigma_1+\sigma_2=6+6$.}
	\label{TwoExtLeg}
\end{table}
We again calculate the combinatorial factors $C_k$ in the standard way by performing the tensor contraction
\begin{align}
&q_{1}^{\mu_1}q_{1}^{\mu_2}\ldots q_{1}^{\mu_{6}}q_{2}^{\mu_7}q_{2}^{\mu_8}\ldots q_{2}^{\mu_{12}}[\eta_{\mathrm{sysm}}^{6}]_{\mu_1\mu_2\ldots\mu_{12}}\nonumber\\
={}&{}720(q_1\cdot q_2)^6+5400(q_1\cdot q_1)(q_1\cdot q_2)^4(q_2\cdot q_2)\nonumber\\
&+4050(q_1\cdot q_1)^2(q_1\cdot q_2)^2(q_2\cdot q_2)^2+225(q_1\cdot q_1)^3(q_2\cdot q_2)^3.\label{expr2}
\end{align}
The combinatorial weight factors $C_{k}$ are the coefficients in \eqref{expr2} which agree with the result obtained by the general formula \eqref{Ck} and are provided in Table \eqref{TwoExtLeg}.\footnote{ A simple observation is that all coefficients must add up to $10395$, i.e. $720+5400+4050+225=10395$. If we had normalized our expression by dividing \eqref{expr2} by $(2\times6-1)!!$, the numerical coefficients of the individual terms would correspond to a particular partition of unity.}

\subsubsection{Three external momenta}
\label{Apppart3}
Finally, we consider the integral \eqref{LogDivInt} for a fixed number of loop momenta ${2\omega=12}$ with three external momenta $q_1$, $q_2$ and $q_{3}$ in the numerator $N(q_1,q_2,q_3)=(q_1\cdot \ell)^{\sigma_1}(q_2\cdot \ell)^{\sigma_2}(q_3\cdot \ell)^{\sigma_3}$ with $2\omega=\sigma_1+\sigma_2+\sigma_3=4+4+4$. The system to be solved is then given by 
\begin{align}
4=&2\sigma_{11}+\sigma_{12}+\sigma_{13},\nonumber\\
4=&2\sigma_{22}+\sigma_{12}+\sigma_{23},\nonumber\\
4=&2\sigma_{33}+\sigma_{13}+\sigma_{23}.
\end{align}
All fifteen non-negative integer solution together with their combinatorial weight factors  are collected in Table \ref{ThreeExtLeg}
\begin{table}[h!]
	\begin{center}
		\begin{tabular}{c|c|c||c|c|c}
			$k$ & $(\sigma_{11},\sigma_{12},\sigma_{13},\sigma_{22},\sigma_{23},\sigma_{33})$&$C_k$&$k$ & $(\sigma_{11},\sigma_{12},\sigma_{13},\sigma_{22},\sigma_{23},\sigma_{33})$&$C_k$\\\hline
			$1$ & $(0,0,4,2,0,0)$ & $72$&
			$9$& $(1,1,1,0,3,0)$ &$1152$\\
			$2$ & $(0,1,3,1,1,0)$ & $1152$&
			$10$& $(1,1,1,1,1,1)$ &$1728$\\
			$3$ & $(0,2,2,0,2,0)$ & $1728 $&
			$11$ & $(1,2,0,0,2,1)$ & $864$\\
			$4$ & $(0,2,2,1,0,1)$ & $864$&
			$12$ & $(1,2,0,1,0,2)$ & $216$\\ 
			$5$ & $(0,3,1,0,1,1)$ & $1152$& 
			$13$ & $(2,0,0,0,4,0)$ & $72$\\
			$6$& $(0,4,0,0,0,2)$ &$72$&
			$14$ & $(2,0,0,1,2,1)$ & $216$\\
			$7$& $(1,0,2,1,2,0)$ &$864 $&
			$15$ & $(2,0,0,2,0,2)$ & $27$\\
			$8$& $(1,0,2,2,0,1)$ &$216$&{}&{}&
		\end{tabular}
	\end{center}
	\caption{All partitions $k$ together with the corresponding values of the $\sigma_{ij}$ and the associated combinatorial weight factor $C_k$ for $2\omega=\sigma_1+\sigma_2+\sigma_3=4+4+4$.}
	\label{ThreeExtLeg}
\end{table}

\noindent We again calculate the combinatorial factors $C_k$ in the standard way by performing the tensor contraction

\begin{align}
&q_{1}^{\mu_1}\ldots q_{1}^{\mu_{4}}q_{2}^{\mu_5}\ldots q_{2}^{\mu_{8}}q_{3}^{\mu_9}\ldots q_{3}^{\mu_{12}}[\eta_{\mathrm{sysm}}^{6}]_{\mu_1\mu_2\ldots \mu_{12}}\nonumber\\
={}&72(q_1\cdot q_3)^4(q_2^2)^2+1152(q_1\cdot q_2) (q_1\cdot q_3)^3q_2^2(q_2\cdot q_3)
\nonumber\\
&+1728(q_1\cdot q_2)^2(q_1\cdot q_3)^2(q_2\cdot q_3)^2+864 q_1^2 (q_1\cdot q_3)^2q_2^2(q_2\cdot q_3)^2\nonumber\\
&+1152 q_1^2 (q_1\cdot q_2)(q_1\cdot q_3) (q_2\cdot q_3)^3+72 (q_1^2)^2(q_2\cdot q_3)^4\nonumber\\
&+864 (q_1\cdot q_2)^2 (q_1\cdot q_3)^2q_2^2 q_3^2+216 q_1^2 (q_1\cdot q_3)^2 (q_2^2)^2 q_3^2\nonumber\\
&+1152 (q_1\cdot q_2)^3 (q_1\cdot q_3)(q_2\cdot q_3) q_3^2\nonumber\\
&+1728 q_1^2 (q_1\cdot q_2)(q_1\cdot q_3) q_2^2 (q_2\cdot q_3) q_3^2\nonumber\\
&+864 q_1^2(q_1\cdot q_2)^2(q_2\cdot q_3)^2 q_3^2+216 (q_1^2)^2 q_2^2 (q_2\cdot q_3)^2 q_3^2\nonumber\\
&+72 (q_1\cdot q_2)^4 (q_3^2)^2+216 q_1^2 (q_1\cdot q_2)^2 q_2^2 (q_3^2)^2\nonumber\\
&+27 (q_1^2)^2(q_2^2)^2(q_3^2)^2.
\end{align}
In general, the number of partitions grows with an increasing number of derivatives $2\omega$ as well as with an increasing number $n$ of different external momenta $q_{i}$, $i=1,\ldots,n$. A physical example for the calculation of the divergent part of a concrete Feynman one-loop integral is provided in the next section for a higher derivative scalar field theory.

\section{Example: higher derivative scalar field theory}
\label{Example}
We consider a one-loop three-point diagram arising in the scalar Galileon theory, which is higher derivative scalar field theory with important applications in cosmology \cite{Nicolis2009,Deffayet2009,Chow2009,Heisenberg2018a}.
Here, we mainly we focus on the off-shell three-point integral for illustrative purposes, as it corresponds to an acceptable balance between the number of generated terms and the demonstration of the efficiency of the combinatorial algorithm. \footnote{From the viewpoint of the on-shell amplitude, the three-point diagrams are trivially zero due to the kinematics: The result can only depend on the ${3(3+1)/2=6}$ invariants $(k_{i}\cdot k_{j})$, $i,j=1,2,3$. Momentum conservation $\sum_{i=1}^{3}k_{i}^{\mu}=0$ reduces the number of independent invariants to $3(3-1)/2=3$. On-shell, the number of independent invariants further reduce by $3$ due to the three conditions $k_{i}^2=0$. Therefore, the number of independent on-shell invariants for the three-point scattering is zero and the result has to vanish on-shell.}

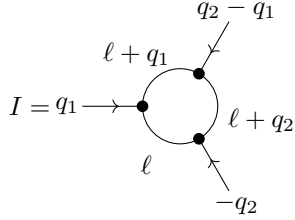
\begin{figure}[h!]
 	\begin{center}
 		\begin{tikzpicture}
 		\begin{scope}[decoration={
 			markings,
 			mark=at position 0.5 with {\arrow{<}}}] 
 		\node at (-2,0) {$I=$};
 		\draw[black] (0,0) circle (0.5cm);
 		\filldraw[black] (60:0.5) circle (0.7mm);
 		\draw[black,postaction={decorate}] (60:0.5) -- (60:1.3);
 		\node at (60:1.5) {$q_{2}-q_{1}$};
 		\node at (130:0.9) {$\ell+q_1$};
 		\node at (350:1.1) {$\ell+q_2$};
 		\node at (240:0.9) {$\ell$};
 		\filldraw[black] (180:0.5) circle (0.7mm);
 		\draw[black,postaction={decorate}] (180:0.5) -- (180:1.3);
 		\node at (180:1.5) {$q_1$};
 		\filldraw[black] (300:0.5) circle (0.7mm);
 		\draw[black,postaction={decorate}] (300:0.5) -- (300:1.3);
 		\node at (300:1.5) {$-q_{2}$};
 		\end{scope}
 		\end{tikzpicture}
 		\caption{Scalar three-point one-loop integral.}
 		\label{ThreePointGalileon}
 	\end{center}
 \end{figure}

In the following, we focus on the particular diagram, shown in Fig. \ref{ThreePointGalileon}.

\noindent
The required Feynman rules for the scalar propagator and the three point vertex that enter the diagram in Fig. \ref{ThreePointGalileon} are collected in Fig. \ref{FRGalileon}. Since the particle in the theory is massless, we introduce an auxiliary IR regulating mass term $m_{\mathrm{IR}}$. The final result for the UV divergences are independent of $m_{\mathrm{IR}}$. 
\begin{figure}[h!]
	\begin{center}
		\begin{tikzpicture}
		\begin{scope}[decoration={
			markings,
			mark=at position 0.5 with {\arrow{>}}}] 
		\node at (3.1,0){$=\;\;\frac{i}{k^2-m^2_{\mathrm{IR}}}$,};
		\node at (1,0.25) {$k$};
		\draw[black,postaction={decorate}](0,0)--(2,0);
		\end{scope}

		\begin{scope}[yshift=-2cm,xshift=0.8cm,decoration={
			markings,
			mark=at position 0.5 with {\arrow{<}}}]
		\filldraw[black] (0:0) circle (0.7mm);
		\draw[black,postaction={decorate}] (60:0) -- (60:0.8);
		\draw[black,postaction={decorate}]  (180:0) -- (180:0.8);
		\draw[black,postaction={decorate}] (300:0) -- (300:0.8);
		\node at (60:1.1){$k_3$};
		\node at (180:1.1){$k_1$};
		\node at (300:1.1){$k_2$};
		\node at (5,0){$=\;\;-4ig_3\left[(k_1\cdot k_2)^2-k_1^2k^2_2\right]+\mathrm{cyclic}(1,2,3)$.};
		\end{scope}
		\end{tikzpicture}
		\caption{Relevant Feynman rules for the scalar Galileon.}
		\label{FRGalileon}
	\end{center}
\end{figure}

\noindent We use the convention that all momenta in a given vertex are incoming. 
and calculate the following integral corresponding to the diagram of Fig. \ref{ThreePointGalileon}:
\begin{align}
I=-1728 g_3^3\int\frac{\mathrm{d}^d\ell}{(2\pi)^4}\frac{N(q_1,q_2;\ell)}{D(q_1,q_2,m_{\mathrm{IR};\ell})}.\label{Gal3PInt}
\end{align}
The external momenta $q_i$ are defined as in \eqref{momq} and we have extracted an overall common numerical factor $-1728$ in the numerator. The numerator and denominator in \eqref{Gal3PInt} are explicitly given by  
\begin{align}
N(q_1,q_2;\ell)={}&\left[(\ell\cdot q_1)^2-\ell^2q_{1}^2\right]\times\left[(\ell\cdot q_2)^2-\ell^2q_{2}^2\right]\nonumber\\
&\times \left\{(\ell\cdot q_1)^2+(\ell\cdot q_2)^2-\ell^2q_1^2-2(\ell\cdot q_2)\left[q_1^2-(q_1\cdot q_2)\right]\right.\nonumber\\
&\left.-2(\ell\cdot q_1)\left[(\ell\cdot q_2)-(q_1\cdot q_2)+q_2^2\right]+2\ell^2(q_1\cdot q_2)\right.\nonumber\\
&\left.+(q_{1}\cdot q_2)^2-\ell^2q_{2}^2-q_1^2q_2^2\right\},\\[1mm]
D(q_1,q_2,m_{\mathrm{IR}};\ell)={}&[\ell^2-m^2_{\mathrm{IR}}]\,[(\ell+q_1)^2-m^2_{\mathrm{IR}}]\,[(\ell+q_2)^2-m^2_{\mathrm{IR}}],
\end{align}
where momentum conservation $q_3=\sum_{i=1}^3k_{i}^{\mu}=0$ has been used in the kinematic parametrization of the integrand in \eqref{Gal3PInt}. 
\paragraph{Step 1:}
Power counting implies that the highest number of loop momenta in the numerator is six. Since the leading power in the loop momentum in the denominator is also six, in $d=4$ dimensions we have to expand all propagators up to to fourth order to extract the logarithmically divergent vacuum integrals. 

For the integral \eqref{Gal3PInt}, the propagator expansion \eqref{PropEx} results in $108$ logarithmically divergent vacuum integrals of the form
\begin{align}
iI^{\mathrm{div}}={}&-1728\, g_3^3\sum_{i=1}^{108}b_i\,I_{i}^{\mathrm{div}},\label{DivInt}\\
I_{i}^{\mathrm{div}}={}&\left.(q_1\cdot q_2)^{\rho_{12}^{i}}(q_1^2)^{\rho_{11}^{i}}(q_2^2)^{\rho_{22}^{i}}\int\frac{\mathrm{d}^4\ell}{(2\pi)^4}\frac{(\ell\cdot q_{1})^{\sigma_{1}^{i}}(\ell\cdot q_2)^{\sigma_{2}^{i}}}{(\ell^2-m_{\mathrm{IR}}^2)^{\lambda^{i}}}\right|^{\mathrm{div}}.
\end{align}
Each of the integrals in the sum can be uniquely characterized by the six numbers $\{i\}:=(\sigma_{1}^{i},\sigma_{2}^{i},\lambda^{i},\rho_{12}^{i},\rho_{11}^{i},\rho_{22}^{i})$, which, together with their numerical coefficients $b_i$ are summarized in Table~\ref{TableDivInt}.\footnote{The numbers are of course not independent: By construction we have $\lambda=2+(\sigma_1+\sigma_2)/2$ and $5-(\sigma_1+\sigma_2)/2=\rho_{12}+\rho_{11}+\rho_{22}$.}
\begin{table}
\begin{center}
\begin{tabular}{c|c|c||c|c|c||c|c|c}
	$i$&$\{i\}$&$b_i$
	&$i$&$\{i\}$&$b_i$
	&$i$&$\{i\}$&$b_i$\\
	\hline
	$1$&$(8, 2, 7, 0, 0, 0)$&$16$
	&$37$&$(6, 2, 6, 0, 0, 1)$&$-4$
	&$73$&$(1, 1, 3, 2, 1, 1)$&$4$\\
	$2$&$(7, 3, 7, 0, 0, 0)$&$-16$
	&$38$&$(3, 5, 6, 0, 0, 1)$&$32$
	&$74$&$(0, 2, 3, 2, 1, 1)$&$5$\\
	$3$&$(3, 7, 7, 0, 0, 0)$&$-16$
	&$39$&$(2, 6, 6, 0, 0, 1)$&$-44$
	&$75$&$(0, 0, 2, 2, 2, 1)$&$-1$\\
	$4$&$(2, 8, 7, 0, 0, 0)$&$16$
	&$40$&$(6, 0, 5, 0, 1, 1)$&$44$
	&$76$&$(6, 0, 5, 0, 0, 2)$&$4$\\
	$5$&$(6, 2, 6, 0, 1, 0)$&$-44$
	&$41$&$(5, 1, 5, 0, 1, 1)$&$-31$
	&$77$&$(4, 2, 5, 0, 0, 2)$&$1$\\
	$6$&$(5, 3, 6, 0, 1, 0)$&$32$
	&$42$&$(4, 2, 5, 0, 1, 1)$&$9$
	&$78$&$(3, 3, 5, 0, 0, 2)$&$-18$\\
	$7$&$(2, 6, 6, 0, 1, 0)$&$-4$
	&$43$&$(3, 3, 5, 0, 1, 1)$&$2$
	&$79$&$(2, 4, 5, 0, 0, 2)$&$41$\\
	$8$&$(1, 7, 6, 0, 1, 0)$&$16$
	&$44$&$(2, 4, 5, 0, 1, 1)$&$9$
	&$80$&$(4, 0, 4, 0, 1, 2)$&$-9$\\
	$9$&$(0, 8, 6, 0, 1, 0)$&$-16$
	&$45$&$(1, 5, 5, 0, 1, 1)$&$-32$
	&$81$&$(3, 1, 4, 0, 1, 2)$&$-2$\\
	$10$&$(4, 2, 5, 0, 2, 0)$&$41$
	&$46$&$(0, 6, 5, 0, 1, 1)$&$44$
	&$82$&$(2, 2, 4, 0, 1, 2)$&$-7$\\
	$11$&$(3, 3, 5, 0, 2, 0)$&$-18$
	&$47$&$(4, 0, 4, 0, 2, 1)$&$-41$
	&$83$&$(1, 3, 4, 0, 1, 2)$&$18$\\
	$12$&$(2, 4, 5, 0, 2, 0)$&$1$
	&$48$&$(3, 1, 4, 0, 2, 1)$&$18$
	&$84$&$(0, 4, 4, 0, 1, 2)$&$-41$\\
	$13$&$(0, 6, 5, 0, 2, 0)$&$4$
	&$49$&$(2, 2, 4, 0, 2, 1)$&$-7$
	&$85$&$(2, 0, 3, 0, 2, 2)$&$6$\\
	$14$&$(2, 2, 4, 0, 3, 0)$&$-14$
	&$50$&$(1, 3, 4, 0, 2, 1)$&$-2$
	&$86$&$(1, 1, 3, 0, 2, 2)$&$2$\\
	$15$&$(1, 3, 4, 0, 3, 0)$&$2$
	&$51$&$(0, 4, 4, 0, 2, 1)$&$-9$
	&$87$&$(0, 2, 3, 0, 2, 2)$&$6$\\
	$16$&$(0, 4, 4, 0, 3, 0)$&$-1$
	&$52$&$(2, 0, 3, 0, 3, 1)$&$14$
	&$88$&$(1, 7, 6, 0, 1, 0)$&$16$\\
	$17$&$(0, 2, 3, 0, 4, 0)$&$1$
	&$53$&$(1, 1, 3, 0, 3, 1)$&$-2$
	&$89$&$(4, 0, 4, 1, 0, 2)$&$4$\\
	$18$&$(6, 2, 6, 1, 0, 0)$&$16$
	&$54$&$(0, 2, 3, 0, 3, 1)$&$2$
	&$90$&$(3, 1, 4, 1, 0, 2)$&$4$\\
	$19$&$(2, 6, 6, 1, 0, 0)$&$16$
	&$55$&$(0, 0, 2, 0, 4, 1)$&$-1$
	&$91$&$(2, 2, 4, 1, 0, 2)$&$18$\\
	$20$&$(4, 2, 5, 1, 1, 0)$&$-32$
	&$56$&$(6, 0, 5, 1, 0, 1)$&$-16$
	&$92$&$(2, 0, 3, 1, 1, 2)$&$-6$\\
	$21$&$(3, 3, 5, 1, 1, 0)$&$-4$
	&$57$&$(4, 2, 5, 1, 0, 1)$&$-4$
	&$93$&$(1, 1, 3, 1, 1, 2)$&$-4$\\
	$22$&$(2, 4, 5, 1, 1, 0)$&$-4$
	&$58$&$(3, 3, 5, 1, 0, 1)$&$-4$
	&$94$&$(0, 2, 3, 1, 1, 2)$&$-18$\\
	$23$&$(0, 6, 5, 1, 1, 0)$&$-16$
	&$59$&$(2, 4, 5, 1, 0, 1)$&$-32$
	&$95$&$(0, 0, 2, 1, 2, 2)$&$2$\\
	$24$&$(2, 2, 4, 1, 2, 0)$&$18$
	&$60$&$(4, 0, 4, 1, 1, 1)$&$32$
	&$96$&$(2, 0, 3, 2, 0, 2)$&$1$\\
	$25$&$(1, 3, 4, 1, 2, 0)$&$4$
	&$61$&$(3, 1, 4, 1, 1, 1)$&$4$
	&$97$&$(2, 4, 5, 0, 0, 2)$&$41$\\
	$26$&$(0, 4, 4, 1, 2, 0)$&$4$
	&$62$&$(2, 2, 4, 1, 1, 1)$&$10$
	&$98$&$(4, 0, 4, 0, 0, 3)$&$-1$\\
	$27$&$(0, 2, 3, 1, 3, 0)$&$-2$
	&$63$&$(1, 3, 4, 1, 1, 1)$&$4$
	&$99$&$(3, 1, 4, 0, 0, 3)$&$2$\\
	$28$&$(4, 2, 5, 2, 0, 0)$&$4$
	&$64$&$(0, 4, 4, 1, 1, 1)$&$32$
	&$100$&$(2, 2, 4, 0, 0, 3)$&$-14$\\
	$29$&$(3, 3, 5, 2, 0, 0)$&$4$
	&$65$&$(2, 0, 3, 1, 2, 1)$&$-18$
	&$101$&$(2, 0, 3, 0, 1, 3)$&$2$\\
	$30$&$(2, 4, 5, 2, 0, 0)$&$4$
	&$66$&$(1, 1, 3, 1, 2, 1)$&$-4$
	&$102$&$(1, 1, 3, 0, 1, 3)$&$-2$\\
	$31$&$(2, 2, 4, 2, 1, 0)$&$-5$
	&$67$&$(0, 2, 3, 1, 2, 1)$&$-6$
	&$103$&$(0, 2, 3, 0, 1, 3)$&$14$\\
	$32$&$(1, 3, 4, 2, 1, 0)$&$-4$
	&$68$&$(0, 0, 2, 1, 3, 1)$&$2$
	&$104$&$(0, 0, 2, 0, 2, 3)$&$-1$\\
	$33$&$(0, 4, 4, 2, 1, 0)$&$-4$
	&$69$&$(4, 0, 4, 2, 0, 1)$&$-4$
	&$105$&$(2, 0, 3, 1, 0, 3)$&$-2$\\
	$34$&$(0, 2, 3, 2, 2, 0)$&$1$
	&$70$&$(3, 1, 4, 2, 0, 1)$&$-4$
	&$106$&$(0, 0, 2, 1, 1, 3)$&$2$\\
	$35$&$(8, 0, 6, 0, 0, 1)$&$-16$
	&$71$&$(2, 2, 4, 2, 0, 1)$&$-5$
	&$107$&$(2, 0, 3, 0, 0, 4)$&$1$\\
	$36$&$(7, 1, 6, 0, 0, 1)$&$16$
	&$72$&$(2, 0, 3, 2, 1, 1)$&$5$
	&$108$&$(0, 0, 2, 0, 1, 4)$&$-1$\\
	\end{tabular}
\caption{Exponents $\{i\}$ of the invariants defining the integrals \eqref{DivInt} together with their corresponding numerical coefficient $b_i$. }
\label{TableDivInt}
\end{center}
\end{table}
\begin{table}
	\begin{tabular}{c|c|c|c||c|c|c|c||c|c|c|c||c|c|c|c}
		$i$&$k$&$\{k\}$&$B_k$
		&$i$&$k$&$\{k\}$&$B_k$
		&$i$&$k$&$\{k\}$&$B_k$
		&$i$&$k$&$\{k\}$&$B_k$\\
		\hline
		$1$&$1$&$(3,2,0)$&$840$
		&$22$&$1$&$(0,2,1)$&$12$
		&$46$&$1$&$(0,0,3)$&$15$
		&$76$&$1$&$(3,0,0)$&$15$\\
		$1$&$2$&$(4,0,1)$&$105$
		&$22$&$2$&$(1,0,2)$&$3$
		&$47$&$1$&$(2,0,0)$&$3$
		&$77$&$1$&$(1,2,0)$&$12$\\
		$2$&$1$&$(2,3,0)$&$630$
		&$23$&$1$&$(0,0,3)$&$15$
		&$48$&$1$&$(1,1,0)$&$3$
		&$77$&$2$&$(2,0,1)$&$3$\\
		$2$&$2$&$(3,1,1)$&$315$
		&$24$&$1$&$(0,2,0)$&$2$
		&$49$&$1$&$(0,2,0)$&$2$
		&$78$&$1$&$(0,3,0)$&$6$\\
		$3$&$1$&$(0,3,2)$&$630$
		&$24$&$2$&$(1,0,1)$&$1$
		&$49$&$2$&$(1,0,1)$&$1$
		&$78$&$2$&$(1,1,1)$&$9$\\
		$3$&$2$&$(1,1,3)$&$315$
		&$25$&$1$&$(0,1,1)$&$3$
		&$50$&$1$&$(0,1,1)$&$3$
		&$79$&$1$&$(0,2,1)$&$12$\\
		$4$&$1$&$(0,2,3)$&$840$
		&$26$&$1$&$(0,0,2)$&$3$
		&$51$&$1$&$(0,0,2)$&$3$
		&$79$&$2$&$(1,0,2)$&$3$\\
		$4$&$2$&$(1,0,4)$&$105$
		&$27$&$1$&$(0,0,1)$&$1$
		&$52$&$1$&$(1,0,0)$&$1$
		&$80$&$1$&$(2,0,0)$&$3$\\
		$5$&$1$&$(2,2,0)$&$90$
		&$28$&$1$&$(1,2,0)$&$12$
		&$53$&$1$&$(0,1,0)$&$1$
		&$81$&$1$&$(1,1,0)$&$3$\\
		$5$&$2$&$(3,0,1)$&$15$
		&$28$&$2$&$(2,0,1)$&$3$
		&$54$&$1$&$(0,0,1)$&$1$
		&$82$&$1$&$(0,2,0)$&$2$\\
		$6$&$1$&$(1,3,0)$&$60$
		&$29$&$1$&$(0,0,0)$&$6$
		&$55$&$1$&$(0,0,0)$&$1$
		&$82$&$2$&$(1,0,1)$&$1$\\
		$6$&$2$&$(2,1,1)$&$45$
		&$29$&$2$&$(1,1,1)$&$9$
		&$56$&$1$&$(3,0,0)$&$15$
		&$83$&$1$&$(0,1,1)$&$3$\\
		$7$&$1$&$(0,2,2)$&$90$
		&$30$&$1$&$(0,2,1)$&$12$
		&$57$&$1$&$(1,2,0)$&$12$
		&$84$&$1$&$(0,0,2)$&$3$\\
		$7$&$2$&$(1,0,3)$&$15$
		&$30$&$2$&$(1,0,2)$&$3$
		&$57$&$2$&$(2,0,1)$&$3$
		&$85$&$1$&$(1,0,0)$&$1$\\
		$8$&$1$&$(0,1,3)$&$105$
		&$31$&$1$&$(0,2,0)$&$2$
		&$58$&$1$&$(0,3,0)$&$6$
		&$86$&$1$&$(0,1,0)$&$1$\\
		$9$&$1$&$(0,0,4)$&$105$
		&$31$&$2$&$(1,0,1)$&$1$
		&$58$&$2$&$(1,1,1)$&$9$
		&$87$&$1$&$(0,0,1)$&$1$\\
		$10$&$1$&$(1,2,0)$&$12$
		&$32$&$1$&$(0,1,1)$&$3$
		&$59$&$1$&$(0,2,1)$&$12$
		&$88$&$1$&$(0,0,0)$&$1$\\
		$10$&$2$&$(2,0,1)$&$3$
		&$33$&$1$&$(0,0,2)$&$3$
		&$59$&$2$&$(1,0,2)$&$3$
		&$89$&$1$&$(2,0,0)$&$3$\\
		$11$&$1$&$(0,3,0)$&$6$
		&$34$&$1$&$(0,0,1)$&$1$
		&$60$&$1$&$(2,0,0)$&$3$
		&$90$&$1$&$(1,1,0)$&$3$\\
		$11$&$2$&$(1,1,1)$&$9$
		&$35$&$1$&$(4,0,0)$&$105$
		&$61$&$1$&$(1,1,0)$&$3$
		&$91$&$1$&$(0,2,1)$&$2$\\
		$12$&$1$&$(0,2,1)$&$12$
		&$36$&$1$&$(3,1,0)$&$105$
		&$62$&$1$&$(0,2,0)$&$2$
		&$91$&$2$&$(1,0,1)$&$1$\\
		$12$&$2$&$(1,0,2)$&$3$
		&$37$&$1$&$(2,2,0)$&$90$
		&$62$&$2$&$(1,0,1)$&$1$
		&$92$&$1$&$(1,0,0)$&$1$\\
		$13$&$1$&$(0,0,3)$&$15$
		&$37$&$2$&$(3,0,1)$&$15$
		&$63$&$1$&$(0,1,1)$&$3$
		&$93$&$1$&$(0,1,0)$&$1$\\
		$14$&$1$&$(0,2,0)$&$2$
		&$38$&$1$&$(0,3,1)$&$60$
		&$64$&$1$&$(0,0,2)$&$3$
		&$94$&$1$&$(0,0,1)$&$1$\\
		$14$&$2$&$(1,0,1)$&$1$
		&$38$&$2$&$(1,2,2,)$&$45$
		&$65$&$1$&$(1,0,0)$&$1$
		&$95$&$1$&$(0,0,0)$&$1$\\
		$15$&$1$&$(0,1,1)$&$3$
		&$39$&$1$&$(0,2,2)$&$90$
		&$66$&$1$&$(0,1,0)$&$1$
		&$96$&$1$&$(1,0,0)$&$1$\\
		$16$&$1$&$(0,0,2)$&$3$
		&$39$&$2$&$(1,0,3)$&$15$
		&$67$&$1$&$(0,0,1)$&$1$
		&$97$&$1$&$(0,0,0)$&$1$\\
		$17$&$1$&$(0,0,1)$&$1$
		&$40$&$1$&$(3,0,0)$&$15$
		&$68$&$1$&$(0,0,0)$&$1$
		&$98$&$1$&$(2,0,0)$&$3$\\
		$18$&$1$&$(2,2,1)$&$90$
		&$41$&$1$&$(2,1,0)$&$15$
		&$69$&$1$&$(2,0,0)$&$3$
		&$99$&$1$&$(1,1,0)$&$3$\\
		$18$&$2$&$(3,0,1)$&$15$
		&$42$&$1$&$(1,2,0)$&$12$
		&$70$&$1$&$(1,1,0)$&$3$
		&$100$&$1$&$(0,2,0)$&$2$\\
		$19$&$1$&$(0,2,2)$&$90$,
		&$42$&$2$&$(2,0,1)$&$3$
		&$71$&$1$&$(0,2,0)$&$2$
		&$100$&$2$&$(1,0,1)$&$1$\\
		$19$&$2$&$(1,0,3)$&$15$
		&$43$&$1$&$(0,3,0)$&$6$
		&$71$&$2$&$(1,0,1)$&$1$
		&$101$&$1$&$(1,0,0)$&$1$\\
		$20$&$1$&$(1,2,0)$&$12$
		&$43$&$2$&$(1,1,1)$&$9$
		&$72$&$1$&$(1,0,0)$&$1$
		&$102$&$1$&$(0,1,0)$&$1$\\
		$20$&$2$&$(2,0,1)$&$3$
		&$44$&$1$&$(0,2,1)$&$12$
		&$73$&$1$&$(0,1,0)$&$1$
		&$103$&$1$&$(0,0,1)$&$1$\\
		$21$&$1$&$(0,3,0)$&$6$
		&$44$&$2$&$(1,0,2)$&$3$
		&$74$&$1$&$(0,0,1)$&$1$
		&$104$&$1$&$(0,0,0)$&$1$\\
		$21$&$2$&$(1,1,1)$&$9$
		&$45$&$2$&$(0,1,2)$&$15$
		&$75$&$1$&$(0,0,0)$&$1$
		&$105$&$1$&$(1,0,0)$&$1$\\
		${}$&${}$&${}$&${}$
		&${}$&${}$&${}$&${}$
		&${}$&${}$&${}$&${}$
		&$106$&$1$&$(0,0,0)$&$1$\\
		${}$&${}$&${}$&${}$
		&${}$&${}$&${}$&${}$
		&${}$&${}$&${}$&${}$
		&$107$&$1$&$(1,0,0)$&$1$\\
		${}$&${}$&${}$&${}$
		&${}$&${}$&${}$&${}$
		&${}$&${}$&${}$&${}$
		&$108$&$1$&$(0,0,0)$&$1$\\
	\end{tabular}
	\caption{All partitions $k\in\mathcal{P}(2\omega;\sigma_{ij})$ for the $i=1,\ldots,108$ logarithmically divergent integrals $I_{i}^{\mathrm{div}}$ together with their combinatorial coefficients $B_k$.}
	\label{PartDiv}
\end{table}

\paragraph{Step 2:} For the evaluation of the loop integrals and the combinatorics, only the two numbers $(\sigma_{1}, \sigma_{2})$ are relevant (i.e. only the first and second entries in the ${i}$ columns of Table \ref{TableDivInt}). For each of the logarithmically divergent integrals, the explicit partitions $\{k\}=(\sigma_{11},\sigma_{12},\sigma_{22})$ of the even integer $2\omega=\sigma_{1}+\sigma_{2}$ into the non-negative integers $\sigma_{11}$, $\sigma_{12}$, $\sigma_{22}$, which correspond to the exponents of the invariants $q_{1}^2$, $(q_1\cdot q_2)$ and $q_2^2$ that can appear in the final result, have to be calculated according to \eqref{SysPart}.
Using the general formula \eqref{LogDivInt2} to extract the divergent part of the actual loop integration, each logarithmically divergent integral $I_{i}^{\mathrm{div}}$ can be written as sum of products of invariants $\mathrm{Inv}_{(i,k)}$ over all partitions $k\in\mathcal{P}(2\omega:\sigma_{ij})$,
\begin{align}
I_{i}^{\mathrm{div}}={}&\frac{i}{(4\pi)^2}\frac{1}{\varepsilon}(q_1\cdot q_2)^{\rho_{12}^{i}}(q_1^2)^{\rho_{11}^{i}}(q_2^2)^{\rho_{22}^{i}}\sum_{k}B_k \mathrm{Inv}_{(i,k)},\label{PartitDivInt}\\
\mathrm{Inv}_{(i,k)}={}&(q_{1}\cdot q_2)^{\sigma_{12}^{(i,k)}}(q_{1}\cdot q_1)^{\sigma_{11}^{(i,k)}}(q_{2}\cdot q_2)^{\sigma_{22}^{(i,k)}}.
\end{align}

\noindent In Table \ref{PartDiv}, we list all partitions for the $108$ logarithmically divergent integrals together with their coefficients $B_k:=C_{k}/P_{\omega}(4)$ defined as the ratio of the combinatorial coefficients $C_{k}$ defined in \eqref{Ck} and the dimensional polynomial defined in \eqref{Pold}.
\paragraph{Step 3:}
Finally, we need to sum over all partitions $k\in \mathcal{P}(2\omega:\sigma_{ij})$ in the divergent integrals \eqref{PartitDivInt} and over all $i=1,\ldots,108$ divergent integrals in \eqref{DivInt} in order to obtain the logarithmically divergent part of the original integral \eqref{Gal3PInt}. The result is
\begin{align}
iI^{\mathrm{div}}={}&-i\frac{432g_{3}^3}{(4\pi)^2\varepsilon}\left[-\frac{1}{3}(q_{1}^2)^3(q_{1}\cdot q_2)^2+(q_1^2)^2(q_1\cdot q_2)^3-\frac{7}{6}q_{1}^2(q_1\cdot q_2)^4+\frac{1}{4}(q_{1}\cdot q_{2})^5\right.\nonumber\\
&\left.-\frac{5}{48}(q_{1}^2)^4q_{2}^2+\frac{3}{4}(q_1^2)^3q_2^2(q_1\cdot q_2)-\frac{23}{12}(q_1^2)^2(q_1\cdot q_2)^2q_2^2+\frac{7}{6}q_{1}^2 (q_{1}\cdot q_2)^3 q_{2}^2\right.\nonumber\\
&\left.-\frac{5}{12}(q_1^2)^3(q_2^2)^2+\frac{37}{80}(q_{1}^2)^2(q_1\cdot q_2)(q_2^2)^2+\left(q_{1}\leftrightarrow q_2\right)\right].\label{OneLoopGal3P}
\end{align}
As required for kinematic reasons, the one-loop divergences \eqref{OneLoopGal3P} trivially vanish on-shell $k_{1}^2=k_{2}^2=0$.  

\bibliography{CombinatoricsFeynmanIntegralsHDScalarTheories}
\end{document}